\documentclass[aps,pre,twocolumn,showpacs,superscriptaddress]{revtex4} % ,twocolumn
\usepackage[T2A]{fontenc}
\usepackage[cp1251]{inputenc}
\usepackage[english]{babel}
\usepackage{graphicx}
\usepackage{color}
\usepackage{bm}
\usepackage{amssymb}
\usepackage[intlimits]{amsmath}
\usepackage{amsmath}

\newcommand{\be}{\begin{equation}}
\newcommand{\ee}{\end{equation}}
\newcommand{\ben}{\begin{equation*}}
\newcommand{\een}{\end{equation*}}

\begin{document}
\title{The $\log\log$ growth of channel capacity for nondispersive nonlinear optical fiber channel in intermediate power range. Extension of the model.}

\author{A.~V.~Reznichenko}
\email[Electronic address: ]{a.v.reznichenko@inp.nsk.su} \affiliation{Budker
Institute of Nuclear Physics of Siberian Branch Russian Academy of Sciences,
Novosibirsk, 630090 Russia} \affiliation{Novosibirsk State University, Novosibirsk,
630090 Russia}
\author{A.~I.~Chernykh}
\email[Electronic address:]{chernykh@iae.nsk.su} \affiliation{Institute of Automation and Electrometry
of the Siberian Branch of the Russian Academy of Sciences,
Novosibirsk, 630090 Russia} \affiliation{Novosibirsk State University, Novosibirsk,
630090 Russia}
\author{S.~V.~Smirnov}
\email[Electronic address:]{smirnov@lab.nsu.ru} \affiliation{Novosibirsk State University, Novosibirsk,
630090 Russia}
\author{I.~S.~Terekhov}
\email[Electronic address:]{i.s.terekhov@gmail.com}\affiliation{Budker
Institute of Nuclear Physics of Siberian Branch Russian Academy of Sciences,
Novosibirsk, 630090 Russia} \affiliation{Novosibirsk State University, Novosibirsk,
630090 Russia}
%\affiliation{Novosibirsk State University, Novosibirsk, 630090 Russia}
%\author{S.~K.~Turitsyn}
%\email[Electronic address:]{s.k.turitsyn@aston.ac.uk} \affiliation{Aston Institute
%of Photonics Technologies, Aston University, Aston Triangle, Birmingham, B4 7ET, UK}
% \affiliation{Novosibirsk State University, Novosibirsk, 630090 Russia}

%=========================================Abstract

%\hskip 2cm
%\date{\today}

\begin{abstract}
In our previous paper \cite{Terekhov:2016a} we considered the optical channel modelled  by the nonlinear Sсhr\"{o}dinger
equation with zero dispersion and  additive Gaussian noise. We found per-sample channel capacity for this model. In the present paper we  extend  per-sample channel model by introducing the initial signal dependence on time and the output signal detection procedure. The proposed model is a closer approximation of the realistic communications link than the per-sample model where there is no dependence of the initial signal on time. For the proposed model we found the correlators of the output signal both analytically and numerically. Using these correlators we built the conditional probability density function. Then we calculated an entropy of the output signal, a conditional entropy, and the  mutual information. Maximizing the mutual information we found the optimal input signal distribution, channel capacity, and their dependence on the shape of the initial signal in the time domain for the intermediate power range.

\end{abstract}
\pacs{89.70.-a, 05.10.Gg} \maketitle

%==================Introduction
\section{Introduction.}

Nonlinear communication channels have received a lot of attention in last twenty years due to the development of fiber optical communication systems.
In these communication systems the Kerr nonlinearity in the optical fiber becomes important when one increases the power of transmitted signal. The problem of capacity finding was considered analytically and numerically in a series of papers, see e.g.
\cite{Mitra:2001,Narimanov:2002,Kahn:2004,Essiambre:2008,Essiambre:2010, Killey:2011, Agrell:2014, Sorokina2014, Terekhov:2016b, Terekhov:2016a, Terekhov:2017} and references therein. In spite of a lot of publications this problem has not been solved for the case of arbitrary Kerr nonlinearity and the second dispersion parameter of an optical fiber. The nondispersive model is much simpler than the case with nonzero dispersion but it catches-up the main features connected with nonlinearity. Also this model is more convenient for the understanding of dependence of the capacity on the channel nonlinearity. So the analytical form of the conditional probability density function  {$P[Y|X]$, i.e., the probability density function (PDF) to receive the output signal $Y$ if the input signal is $X$},  for nondispersive per-sample channel was first obtained in Ref.\cite{M1994}. The upper bound for the capacity at very large input signal power for the model was obtained in Refs. \cite{tdyt03,Mansoor:2011}. The capacity of the channel was found in Ref.~\cite{Terekhov:2016a} in the intermediate power range implying both large signal-to-noise ratio (SNR) and the condition for the next-to-leading  corrections in the noise power to be small, see Eq.~(23) in Ref.~\cite{Terekhov:2016a}.

The per-sample model assumes that the input signal does not depend on time. In realistic communication channel the transmitted signal does depend on time. In the recent paper \cite{Kramer:2017}  the influence of the receiver, signal, and noise bandwidth on the autocorrelation function and the capacity  was discussed  within the filter-and-sample model for the channel with zero dispersion. In our opinion, one of the important result of the paper \cite{Kramer:2017} is understanding that the conditional PDF depends significantly on the properties of the receiver.

In this paper we consider the nondispersive channel {in the intermediate power range in the case} where the initial signal depends on time and has the bandwidth much less than the noise bandwidth. We also introduce the detection procedure which takes into account the time resolution characteristics of the detector and we demonstrate the influence of the detector and the noise bandwidth on statistical properties of the channel. Therefore  this paper is the generalization of the previous results of Refs.\cite{Terekhov:2016a,Terekhov:2017} for the per-sample model to the time-dependant signal.

The paper is organized in the following way. In Sec. II we present the model of the signal propagation, the input signal  and the receiver  model. In Sec. III  we obtain the conditional probability density function for the introduced model. In Sec. IV we present numerical results for the correlators and compare these results with analytical ones. And in Sec. V we  calculate the optimal input signal distribution and the channel capacity in the intermediate power range. In the {Conclusion} we discuss our results.

%=========================================
\section{Model of the signal propagation and detection }

%===============Model of the signal propagation
In our model the propagation of the signal $\psi(z,t)$  is described by the stochastic nonlinear Sсhr\"{o}dinger equation (NLSE) {with zero dispersion}:
\begin{eqnarray}\label{startingCannelEqt}
&&\!\!\!\partial_z \psi-i\gamma |\psi|^2 \psi=\eta(z,t) \,,
\end{eqnarray}
where  $\gamma$ is the Kerr nonlinearity
coefficient, the function $\psi(z,t)$ obeys the
input and output  conditions:  $\psi(z=0,t)=X(t)$ and $\psi(z=L,t)=Y(t)$, respectively,  $L$ is the length of the signal propagation, and
$\eta(z,t)$ is an additive complex  noise with zero mean $\langle
\eta (z,t)\rangle_{\eta}=0$, and the correlation function in the frequency domain:
\begin{eqnarray}\label{noisecorrelatorw}
\langle \eta (z,\omega)\bar{\eta}(z^\prime,\omega^\prime)\rangle_{\eta} &=&  2\pi Q \delta(\omega-\omega^\prime) \theta\left(\frac{W^\prime}{2}-|\omega|\right)\times  \nonumber \\&&
\delta(z-z^\prime)  \,,
\end{eqnarray}
where the bar means complex conjugation, and $Q$ is a power of the noise per unit length and per unit frequency, $\theta(\omega)$ is the Heaviside theta-function,  $\delta(\omega)$ is the Dirac delta-function, $W'$ is the bandwidth of the noise. {The noise $\eta (z,\omega)$ is not white due to limited bandwidth.} In the time domain this correlator has the form
\begin{eqnarray}\label{noisecorrelatort}
\langle \eta (z,t)\bar{\eta}(z^\prime,t^\prime)\rangle_{\eta} =  Q \frac{W'}{2\pi}\mathrm{sinc}\left(\frac{W' (t-t^\prime)}{2} \right)
\delta(z-z^\prime)\,.
\end{eqnarray}
One can see that if the time difference $t-t'=2 n \pi/W'$ then the correlator (\ref{noisecorrelatort}) is equal to zero, here $n$ is integer.  Thus we can solve equation (\ref{startingCannelEqt}) independently for parameters $t_j=j \Delta$  for different integer $j$, where $\Delta=2\pi/W'$ is the time grid spacing. Therefore instead of the continuous time model (\ref{startingCannelEqt}) we will consider the following discrete model:
\begin{eqnarray}\label{startingCannelEqtDM}
&&\!\!\!\partial_z \psi(z,t_j)-i\gamma |\psi(z,t_j)|^2 \psi(z,t_j)=\eta(z,t_j) \,
\end{eqnarray}
for any time moment $t_j$. It means that we obtain the set of independent time channels since the noise in these moments is not correlated. 
We present the input and output conditions in the discrete form as well: $\psi(z=0,t_j)=X(t_j)$ and $\psi(z=L,t_j)=Y(t_j)$.
Note that the solution $\Phi(z,t_j)$ of the equation (\ref{startingCannelEqtDM}) with zero noise which obeys the input condition $\Phi(z=0,t_j)=X(t_j)$ has the form:
\begin{eqnarray}\label{Phifunction}
&&\Phi(z,t_j)=X(t_j)e^{i \gamma z |X(t_j)|^2}.
\end{eqnarray}
Below we assume that the frequency bandwidth $W'$ of the noise is much broader than the frequency bandwidth $W$ of the input signal $X(t)$  and the frequency bandwidth $\widetilde{W}$ of the function $\Phi(z=L,t)$.
%In this case one can treat the noise as a white Gaussian one, since we have the formal limit
%\begin{eqnarray}\label{limit}
%&& \lim_{W' \to \infty} \frac{W'}{2\pi}\mathrm{sinc}\left({W' (t-t^\prime)}/{2} \right)= \delta (t-t^\prime).
%\end{eqnarray}

%===============Model of the initial signal
In our model the input signal $X(t)$ has the form:
\begin{eqnarray}\label{Xtmodelg}
X(t)=\sum^{N}_{k=-N} C_{k} \, f(t-k T_0),
\end{eqnarray}
where $C_{k}$ are complex random coefficients with some probability density function $P_X[\{C\}]$, $\{C\}=\{C_{-N},\ldots,C_{N} \}$; the pulse envelope $f(t)$ is
the real function which is normalized as $\int^{\infty}_{-\infty}\dfrac{dt}{T_0} f^2(t)=1$. The pulse envelope $f(t)$ has the following properties:  the overlapping of the functions $f(t-k T_0)$ and $f(t-m T_0)$ for $k\neq m$ is  {negligible}: $\int^{\infty}_{-\infty} dt f(t-k T_0) f(t-m T_0) \approx 0$. It means that the function $f(t)$ has  {almost} the finite support $[-T_0/2, T_0/2]$, and the input signal $X(t)$ is defined on the interval $T=(2N+1)T_0$. Thus the frequency support of the function $X(t)$ is infinite. But we imply that  {$\int_{W} |X(\omega)|^2 d\omega \approx \int_{W'} |X(\omega)|^2 d\omega$}, where $X(\omega)$ is the Fourier transformation of $X(t)$. The last relation means that $T_0 W \gg 1$.

In our consideration the average input signal power $P$ is fixed:
\begin{eqnarray}\label{avpower}
P= \int \left(\prod_{k=-N}^{N} d^2 C_k\right)
P_X[\{C\}]\int^{\infty}_{-\infty}\frac{dt}{T} |X(t)|^2,
\end{eqnarray}
where $d^2 C_k= d \mathrm{Re} C_k d \mathrm{Im} C_k$, and the input signal probability density function $P_X[\{C\}]$ is normalized as follows:
\begin{eqnarray}\label{normalization0}
\int \left(\prod_{k=-N}^{N} d^2 C_k\right) P_X[\{C\}]=1.
\end{eqnarray}
Using the properties of the function $f(t-k T_0)$ we can rewrite equation (\ref{avpower}):
\begin{eqnarray}\label{avpower-1}
P=\int  d^2 C_m  P^{(m)}_X[C_m] |C_{m}|^2,
\end{eqnarray}
where
\begin{eqnarray}
P^{(m)}_X[C_m]= \int \left(\prod_{k=-N, k \ne m}^{N} d^2 C_k\right) P_X[\{C\}],
\end{eqnarray}
and we imply that the distribution $P^{(m)}_X[C_m]$ does not depend on $m$.

Let us describe the output signal detection procedure. Our detector recovers the information which is carried by the coefficients $\{C_k\}$. First, the detector receives the signal $\psi(z=L,t_j)$  at the discrete time moments $t_j=j \Delta$, here $j=-M, \ldots, M-1$, where $M=T/(2\Delta) \gg N$.
It means that the time resolution of the detector coincides with the time discretization $\Delta$. Since $\Delta \ll 1/\widetilde{W}$ our detector can completely recover the input signal in noiseless case. Second, the detector removes the nonlinear phase to obtain the recovered input signal $\widetilde{X}(t)$ in the following form
\begin{eqnarray}\label{tildeXt}
\widetilde{X}(t_j)=\psi(z=L,t_j)e^{-i \gamma L |\psi(z=L,t_j)|^2}.
\end{eqnarray}
And finally, using $\widetilde{X}(t)$ detector recovers the coefficients $\widetilde{C}_k$ by projecting on the basis functions $f(t-k T_0)$:
\begin{eqnarray}\label{tildeCk}
\widetilde{C}_k&=&\frac{1}{T_0}\int^{\infty}_{-\infty} dt f(t-k T_0)\widetilde{X}(t)  \nonumber \\
&\approx& \frac{\Delta}{T_0} \sum_{j=-M}^{M-1} f(t_j-k T_0)\widetilde{X}(t_j).
\end{eqnarray}
One can check that in the case of zero noise $\widetilde{X}(t)=X(t)$ and $\widetilde{C}_k=C_k$.

%==========================================C_k
\section{Statistics of $\widetilde{C}_k$}

In the previous paper \cite{Terekhov:2016a} we obtained the conditional probability function $P[Y|X]$ for the case where input $X$ and output
$Y$ signals do not depend on time (per-sample conditional PDF). In the previous section we extend our  model \cite{Terekhov:2016a} by including detector procedure and time dependence of the input signal $X(t)$. Our goal is to obtain conditional probability function  $P[\{\widetilde{C}\}|\{C\}]$, i.e., the probability to detect the set of coefficients $\{\widetilde{C}\}$ if the transmitted set is $\{{C}\}$. Using the function $P[\{\widetilde{C}\}|\{C\}]$ we can calculate the probability density function $P_{out}[\{\widetilde{C}\}]$ as
\begin{eqnarray}\label{PoutCk}
P_{out}[\{\widetilde{C}\}]=\int \prod^{N}_{k=-N} d^2 C_k P[\{\widetilde{C}\}|\{C\}] P_X[\{C\}].
\end{eqnarray}

Since the propagation of the signal in the different time moments $t_j$ is independent, and noise is not correlated, 
the conditional probability function $P[Y(t)|X(t)]$, i.e., the probability density to obtain the output signal $Y(t)$ for the given  input signal $X(t)$, can be presented in the factorized form:
\begin{eqnarray}\label{PYXproduct}
P[Y(t)|X(t)]=\prod^{M-1}_{j=-M} P_j[Y_j|X_j],
\end{eqnarray}
where $X_j=X(t_j)$, $Y_j=Y(t_j)$, and $P_j[Y_j|X_j]$ is per-sample conditional PDF obtained in Ref.~\cite{Terekhov:2016a}.
The  function $P_j[Y_j|X_j]$ in the leading and next-to-leading order in parameter $\sqrt{Q}$ can  {be deduced from the results of Ref.~\cite{Terekhov:2016a}}, where we have to replace parameter $Q$ by $Q/\Delta$:
\begin{widetext}
\begin{eqnarray}\label{PYXapprox}
&&\!\!\!P_j[Y_j|X_j]=\Delta \frac{\exp\left\{- \Delta\dfrac{(1+4\mu_{(j)}^2/3)x^2_{(j)}-2\mu_{(j)} x_{(j)} y_{(j)}+y^2_{(j)}}{{Q} L
(1+\mu_{(j)}^2/3)}\right\}} {\pi {Q} L \sqrt{1+\mu_{(j)}^2/3}}\Bigg(1- \frac{\mu_{(j)}/\rho_{(j)}}{15
(1+\mu_{(j)}^2/3)^2}\Big(\mu_{(j)} (15+\mu_{(j)}^2)x_{(j)}-\nonumber \\&& 2(5-\mu_{(j)}^2/3)y_{(j)}\Big)-
\frac{\mu_{(j)}\Delta}{135 {Q} L \rho_{(j)}\left(1+\mu_{(j)} ^2/3\right)^3 }\Big\{ \mu_{(j)} \left(4 \mu_{(j)} ^4+15 \mu_{(j)}
^2+225\right) x^3_{(j)}+ \left(23 \mu_{(j)} ^4+255 \mu_{(j)} ^2-90\right) x^2_{(j)} y_{(j)}+ \nonumber \\&& \mu_{(j)} \left(20
\mu_{(j)} ^4+117 \mu_{(j)} ^2-45\right) x_{(j)} y^2_{(j)}-  3 \left(5 \mu_{(j)} ^4+33 \mu_{(j)}
^2+30\right) y^3_{(j)} \Big\}\Bigg),
\end{eqnarray}
\end{widetext}
here  $\rho_{(j)}= |X_j|$, $X_j=\rho_{(j)} e^{i \phi_{(j)}} $,  $\mu_{(j)}=\gamma L \rho_{(j)}^2 $, and $x_{(j)}= \mathrm{Re}\left[Y_j e^{-i \phi_{(j)}-i\mu_{(j)}}-\rho_{(j)} \right]$,  $y_{(j)}=\mathrm{Im}\left[Y_j e^{-i \phi_{(j)}-i\mu_{(j)}}-\rho_{(j)}\right]$. The expression (\ref{PYXapprox}) was obtained in Ref.~\cite{Terekhov:2016a} on the condition that the average input signal power $P$ lies in the intermediate power range:
\begin{eqnarray}\label{region}
\frac{Q L}{\Delta} \ll P \ll \Delta /\left( Q L^3 \gamma^2 \right),
\end{eqnarray}
where $P=2\pi \int^{\infty}_{0} d\rho \rho^3 P[\rho]$, $P[\rho]$ is
the distribution function of the quantity $\rho$, see
Ref.~\cite{Terekhov:2016a}. Therefore, our consideration is restricted by the condition (\ref{region}).
The factorization of $P[Y(t)|X(t)]$ in
the form (\ref{PYXproduct}) means that there are $2M$ independent
``sub-channels''. Note that, the signal $X(t)$ is completely defined
by $2N+1$ coefficients $C_k$, i.e., there are only $2N+1$
independent $X_j$, but all $2M$ quantities $Y_j$ are independent.
However, our detector reduces the function $Y(t)$ to the set of
$2N+1$ coefficients $\{\widetilde{C}_k\}$ by the procedure
(\ref{tildeXt}) and (\ref{tildeCk}). Therefore we have to reduce the
function $P[Y(t)|X(t)]$ to the function
$P[\{\widetilde{C}_k\}|\{C_k\}]$ by integrating  over $2M-2N-1$
redundant degrees of freedom. Using the conditional PDF
$P[Y(t)|X(t)]$ in the form (\ref{PYXproduct}) one can calculate all
correlators of the coefficients $\widetilde{C}_{k}$: $\langle
\widetilde{C}_{k_1} \rangle$, $\langle \widetilde{C}_{k_1}
\widetilde{C}_{k_2}\rangle$, $\langle \widetilde{C}_{k_1} \ldots
\widetilde{C}_{k_n}\rangle$. Here
\begin{eqnarray}\label{CorrelatorCC}
&& \langle \widetilde{C}_{k_1} \ldots \widetilde{C}_{k_n}\rangle =\nonumber \\&& \int \prod^{M-1}_{j=-M} d^2Y_j P[Y(t)|X(t)] \widetilde{C}_{k_1} \ldots \widetilde{C}_{k_n},
\end{eqnarray}
where $d^2Y_j= d \mathrm{Re} Y_j d \mathrm{Im} Y_j$, and $\widetilde{C}_{k}$ is defined in equation (\ref{tildeCk}), and in the discrete form it reads:
\begin{eqnarray}\label{Ctildedef}
\widetilde{C}_{k}=\frac{\Delta}{T_0}\sum^{M-1}_{j=-M} f(t_j-k T_0) Y_j e^{-i \gamma L |Y_j|^2}.
\end{eqnarray}
To recover the function $P[\{\widetilde{C}_k\}|\{C_k\}]$ in the leading  approximation in parameter $Q$ it is necessary to know only three correlators:
$\langle \widetilde{C}_k \rangle$, $\langle \widetilde{C}_k \widetilde{C}_m \rangle$,  $\langle \widetilde{C}_k \overline{\widetilde{C}}_m \rangle$.
After substitution of  Eqs. (\ref{PYXproduct}), (\ref{PYXapprox}), and~(\ref{Ctildedef}) to Eq.~(\ref{CorrelatorCC}) and performing the integration we obtain {in the leading order in the noise parameter Q}:
\begin{eqnarray}\label{avC}
&&\langle \widetilde{C}_k \rangle= C_k- \frac{i{C_k\, {Q} L^2  \gamma}}{\Delta} \left(1 - \frac{i\gamma L |C_k|^2 n_4}{3} \right),
\end{eqnarray}
\begin{eqnarray}\label{avCC}
&&\left\langle \left(\widetilde{C}_m - \langle \widetilde{C}_m \rangle \right)\left(\widetilde{C}_n - \langle \widetilde{C}_n \rangle \right)  \right\rangle= \nonumber \\&& -i \delta_{m,n} \frac{ {C}^2_m  {Q} L^2 \gamma}{T_0}\left(n_4-\frac{2i n_6}{3}\gamma L |{C}_m|^2\right) ,
\end{eqnarray}
\begin{eqnarray}\label{avCbarC}
&&\left\langle \left(\widetilde{C}_m - \langle \widetilde{C}_m \rangle \right)\overline{\left(\widetilde{C}_n - \langle \widetilde{C}_n \rangle \right)}  \right\rangle= \nonumber \\&& \delta_{m,n} \frac{{Q} L}{T_0} \left(1+\frac{2 n_6}{3}\gamma^2 L^2 |{C}_m|^4\right) ,
\end{eqnarray}
where $\delta_{m,n}$ is Kronecker symbol and
\begin{eqnarray}\label{ns}
n_s=\int^{T_0/2}_{-T_0/2}\dfrac{dt}{T_0} f^s(t).
\end{eqnarray}
Note that for the first correlator $\left(\langle \widetilde{C}_k- {C}_k \rangle\right)$ is proportional to $Q L/\Delta= Q L W'/(2\pi)$, i.e.,
it is proportional to the total noise power. Whereas the correlators (\ref{avCC}) and (\ref{avCbarC}) are proportional to $Q L/T_0$ and do not depend on the discretization parameter $\Delta$ only in leading order in parameter $Q$ and depend on the parameter $\Delta$ in higher order corrections in
parameter $Q$, see Appendix \ref{AppendixA}.

%We also perform the numerical simulations and compare them with analytical results (\ref{avC})--(\ref{avCbarC}), see fig.~1a,1b,1c. The details of the numerical simulations are presented in the Appendix.

Using the correlators (\ref{avC})--(\ref{avCbarC}) we obtain the conditional PDF $P[\widetilde{C}|C]$ in the leading order in parameter ${Q}$:
\begin{eqnarray}\label{PCC}
P[\widetilde{C}|C]=\prod^{N}_{m=-N}P_{m}[\widetilde{C}_m|C_m],
\end{eqnarray}
where
\begin{widetext}
\begin{eqnarray}\label{PmCC}
P_{m}[\widetilde{C}_m|C_m] \approx  \frac{T_0}{\pi Q L \sqrt{1+\xi^2
\mu^2_{m}/3}} \exp\left[- T_0  \frac{\left(1+4n_6 \mu^2_m
/(3)\right)x^2_m+2 x_m y_m \mu_m n_4  +y^2_m}{Q L \left(1+\xi^2
\mu^2_{m}/3\right)}\right].
\end{eqnarray}
\end{widetext}
Here we have introduced the  notations:
\begin{eqnarray}
x_m&=& \mathrm{Re} \Bigg[e^{-i \phi_m} \Big\{\widetilde{C}_m-{C}_m+\nonumber\\&&\frac{i{C_m \gamma\, {Q} L}}{T_0} \Big(1 - \frac{i\gamma L |C_m|^2 n_4}{3} \Big)\Big\}\Bigg],
\end{eqnarray}
\begin{eqnarray}
y_m&=& \mathrm{Im} \Bigg[e^{-i \phi_m} \Big\{\widetilde{C}_m-{C}_m+\nonumber\\&&\frac{i{C_m \gamma\, {Q} L}}{T_0} \Big(1 - \frac{i\gamma L |C_m|^2 n_4}{3} \Big)\Big\}\Bigg], \\
\phi_m&=&\arg {C}_m, \qquad \mu_m=\gamma L |C_m|^2, \\ \label{xi} \xi^2&=&(4n_6 - 3 n^2_4).
\end{eqnarray}
The parameter $\xi^2$ obeys inequality  $\xi^2>n_6 >0$ due to Cauchy-Schwarz-Buniakowski inequality.
Note that the function $P[\widetilde{C}|C]$ has the factorized form (\ref{PCC}) only in the leading approximation in the parameter $Q$.
The Eq.~(\ref{PCC}) means that we have $2N+1$ independent information channels, and the channel corresponding to the time slot $m$ is described by the function  $P_{m}[\widetilde{C}_m|C_m]$. The function $P_{m}[\widetilde{C}_m|C_m]$ obeys the normalization condition
\begin{eqnarray}
\int d^2 \widetilde{C}_m P_{m}[\widetilde{C}_m|C_m]=1.
\end{eqnarray}
Since there are $2N+1$ independent channels,  we can choose the input signal distribution $P_X[\{C_m\}]$ in the factorized form:
\begin{eqnarray}
P_X[\{C\}]=\prod^N_{k=-N} P^{(k)}_X[C_k],
\end{eqnarray}
and we can consider only one channel, say $m$-th channel. For this channel we can calculate  the probability distribution function of the coefficients $\widetilde{C}_m$:
\begin{eqnarray}\label{Poutm1}
P^{(m)}_{out}[\widetilde{C}_m]= \int d^2 {C}_m  P_{m}[\widetilde{C}_m|C_m]  P^{(m)}_{X}[C_m].
\end{eqnarray}
We imply that the function $P^{(m)}_{X}[C_m]$ is a smooth function
that changes on a scale $|C_m|^2 \sim P $ which is much greater than
$Q L/\Delta$:
\begin{eqnarray}
P \gg  Q L/\Delta \gg Q L/T_0.
\end{eqnarray}
In other words, the signal power is much greater than the noise power in the channel.
The variation scale of the function $ P_{m}[\widetilde{C}_m|C_m]$ in the variable $C_m$ is of order of $\sqrt{Q L/T_0}$ therefore we can use
Laplace's method \cite{Lavrentiev:1987} for the calculation of the integral (\ref{Poutm1}). Performing the integration in the leading order in parameter $Q$ we obtain
\begin{eqnarray}\label{Poutm2}
P^{(m)}_{out}[\widetilde{C}_m]\approx P^{(m)}_{X}[\widetilde{C}_m],
\end{eqnarray}
for details see Appendix C in Ref.~\cite{Terekhov:2016a}. The result (\ref{Poutm2}) implicates that the statistics of the coefficients $\widetilde{C}_m$ coincides with the statistics of the coefficients $C_m$.

%==========================================Numerical calculations of the correlators
\section{Numerical calculations of the correlators}

In order to verify analytical results we  performed numerical simulations
of pulse propagation through nonlinear nondispersive optical fiber and calculated correlators (\ref{avC}), (\ref{avCC}), and (\ref{avCbarC}).
For these purposes we solve numerically Eq.~(\ref{startingCannelEqt}) for fixed input signal $X(t)$ and for different realizations of the noise $\eta(z,t)$. Then we numerically perform the detection procedure described by Eqs.~ (\ref{tildeXt}), (\ref{tildeCk}). Finally, we average the coefficients $\widetilde{C}_k$ and their quadratic combinations over noise realizations. In our simulations we use two numerical methods of the solution of  Eq.~(\ref{startingCannelEqt}): the split-step Fourier method and Runge-Kutta method of the fourth order. The results are presented in the following subsections. We have checked that  the  numerical results do not depend on the numerical method and these results are consistent with analytical ones for different realizations of the form $f(t)$ of the input pulse.

For numerical simulation we choose the following realistic channel parameters. The duration of one pulse is $T_0= 10^{-10}$~sec; fiber length is equal to $L=800$~km; Kerr nonlinearity parameter is  $\gamma=1.25$~(km$\times$W)$^{-1}$.

%=======================================================
\subsection{Split-step Fourier method}

Equation (\ref{startingCannelEqt}) was integrated numerically over $z$ from 0 up to
communication line length $L$ using split-step Fourier method \cite{Agraval:2007}, \cite{Hardin:1973}:
\begin{eqnarray}\label{eqNumStep}
\!\!\! \psi(z+h,t) = \psi(z,t) \exp\left(i\gamma|\psi(z,t)|^2 h\right) +     \hat{F}_{-} \{ \delta Q_h \},
\end{eqnarray}
where $\psi(z,t)$ stands for numerical solution of (\ref{startingCannelEqt}),
$h$ is a step size of $z$-mesh,
$\hat{F}_{-}$ denotes discrete inverse Fourier transform. The quantity $\delta Q_h$ stands for the noise addition per step $h$ which is made in frequency domain according to
\begin{equation}\label{eqNoiseAddition}
  \psi(z,\omega_j) \rightarrow \psi(z,\omega_j) +
    \sqrt{\frac{h Q}{T}} \cdot
    \frac{\eta_X + i\eta_Y}{\sqrt{2}},
\end{equation}
where $j = 0, \ldots, 2M-1$  stands for index of $\omega$-mesh,
$2M$ is the number of $t$- and $\omega$-mesh points, $T$ is the
total width of $t$-mesh, we choose $T=64 T_0$, see
Eq.~(\ref{Xtmodelg2}) below; $\eta_X$ and $\eta_Y$ are independent
standard Gauss random numbers with zero mean and $\sigma^2=1$,
additive noise level is $Q=10^{-21}$~W/(km$\times$Hz).

The input signal for $z=0$ has the form
\begin{eqnarray}\label{Xtmodelg2}
\psi(z=0,t) =X(t)=\sum^{64}_{k=1} C_{k} \, f(t-k T_0),
\end{eqnarray}
here we use the pulse envelope of the Gaussian form:
\begin{equation}\label{eqInitCond}
  f(t) =
    \sqrt{\frac{T_0}{T_1\sqrt{\pi}}}
    \exp\left( -\frac{t^2}{2T_1^2} \right),
\end{equation}
where  $T_1=T_0/10=10^{-11}$~sec stands for the characteristic time
scale of the function $ f(t)$.  {Pulse
intersection is negligible.} For such pulses coefficients $n_s$
defined in Eq.~(\ref{ns}) are $n_4 = \frac{T_0/T_1}{\sqrt{2\pi}}
\approx 3.989$, $n_6 = \frac{(T_0/T_1)^2}{\pi\sqrt{3}} \approx
18.38$, $n_8 = \frac{(T_0/T_1)^3}{2\pi\sqrt{\pi}} \approx 89.79$,
$\xi \approx 5.08$.

In the numerical simulation we  vary the  average power
$\frac{1}{64}\sum^{64}_{k=1} |C_k|^2$ of the input signal from
$0.0177$~mW up to $4.43$~mW.  It corresponds to the variation of the
peak power ($|C_k|^2 f^2(0)$)  from $0.1$~mW up to 25~mW.

Simulations are performed  for different $t$-meshes (different grid spacing $\Delta$), i.e., for different noise bandwidths and fixed noise parameter $Q$. 
These meshes differ from each other by time grid spacing $\Delta=T/(2M)$:
$ \Delta_1= 9.77\times 10^{-14}$~sec, $\Delta_2=1.95\times
10^{-13}$~sec and $\Delta_3=3.91\times 10^{-13}$~sec. These grid
spacings determine the widths of conjugated $\omega$-meshes:
$1/\Delta_1=10.26$~THz, $1/\Delta_2=5.12$~THz and
$1/\Delta_3=2.56$~THz.

For each average power of the signal and each mesh step  size
$\Delta$ we simulate propagation of the signal for different
realizations of the noise and then average obtained results for
correlators over realizations. The total number of  noise
realizations for fixed $X(t)$, see Eq. (\ref{Xtmodelg2}), is
determined by the necessary statistic relative error and is chosen
as $5.0 \times 10^4$. This number of the realizations corresponds
to the statistic relative error for  correlators (\ref{avCC}) and
(\ref{avCbarC})  on the level of $0.2\%$ (since the total number of
pulses is $64 \times 5.0 \times 10^4=3.2\times 10^6$). We performed
simulations on $z$-meshes with different number of points (100, 200,
400, 800) and checked out that the results do not depend on step
size $h$.

%========FIGURE-1
\begin{figure}[h]
\begin{center}
\includegraphics[width=6.9cm]{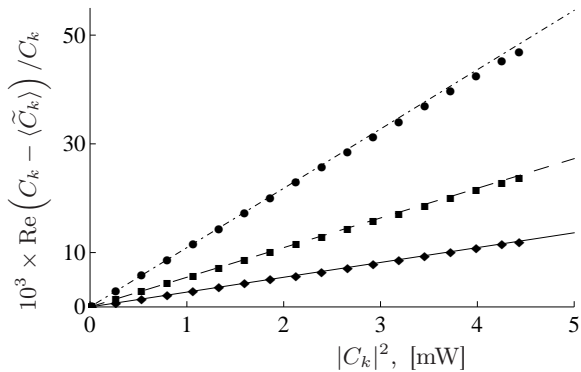}
\begin{picture}(0,0)
\put(-96,-10){\text{$|C_k|^2$, \,[mW]}}
\put(-220,10){\rotatebox{90}{$10^3\times \mathrm{Re} \left({C}_k- \langle \widetilde{C}_k \rangle  \right)/{C}_k $}}
\end{picture}
\end{center}
\caption{\label{figure1} The real part of the relative difference of  the coefficient $C_k$ and the correlator (\ref{avC}) in units $10^{-3}$
as a function of input signal power $|C_k|^2$ for $f(t)$ from Eq.~(\ref{eqInitCond}). The noise power parameter is $Q=10^{-21}$~W/(km$\times$Hz).
Dashed doted, dashed, and solid lines correspond to analytic representation (\ref{avC}) for time grid spacings $\Delta_1$, $\Delta_2$, $\Delta_3$, respectively. Circles, squares, and diamonds correspond to numerical results for time grid spacings $\Delta_1$, $\Delta_2$, $\Delta_3$, respectively.}
\end{figure}
%========FIGURE-2
\begin{figure}[h]
\begin{center}
\includegraphics[width=6.9cm]{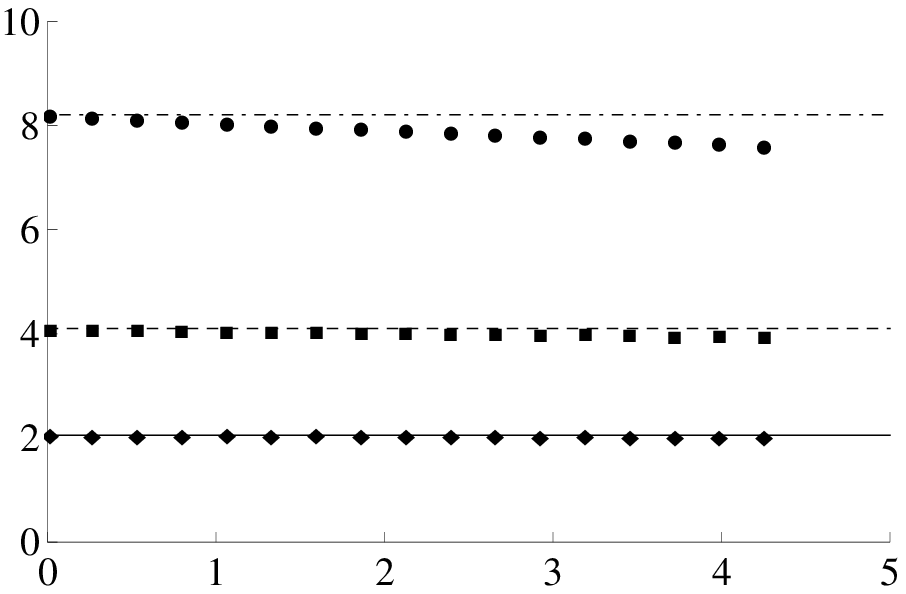}
\begin{picture}(0,0)
\put(-96,-10){\text{$|C_k|^2$, \,[mW]}}
\put(-220,10){\rotatebox{90}{$10^3\times \mathrm{Im} \left({C}_k- \langle \widetilde{C}_k \rangle  \right)/{C}_k $}}
\end{picture}
\end{center}
\caption{\label{figure2} The imaginary part of the relative difference of  the coefficient $C_k$ and the correlator (\ref{avC}) in units $10^{-3}$
as a function of input signal power $|C_k|^2$ for $f(t)$ from Eq.~(\ref{eqInitCond}). The noise power parameter is $Q=10^{-21}$~W/(km$\times$Hz).
Dashed doted, dashed, and solid lines correspond to analytic representation (\ref{avC}) for time grid spacings $\Delta_1$, $\Delta_2$, $\Delta_3$, respectively. Circles, squares, and diamonds correspond to numerical results for time grid spacings $\Delta_1$, $\Delta_2$, $\Delta_3$, respectively.}
\end{figure}
%========FIGURE-3
\begin{figure}[h]
\begin{center}
\includegraphics[width=6.9cm]{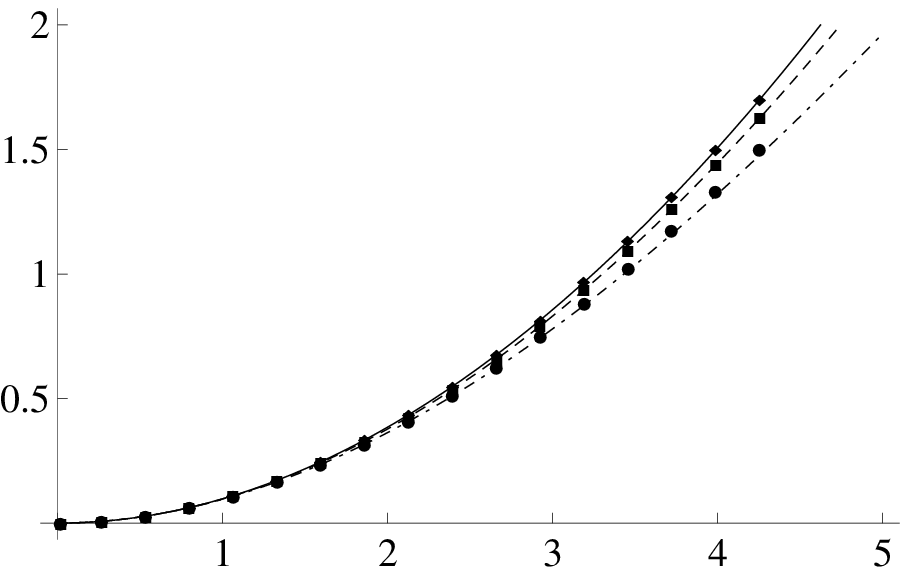}
\begin{picture}(0,0)
\put(-96,-10){\text{$|C_m|^2$, \,[mW]}}
\put(-220,10){\rotatebox{90}{$-\mathrm{Re}\langle \left(\widetilde{C}_m - \langle \widetilde{C}_m \rangle \right)^2 \rangle   , \,[\mu W]$}}
\end{picture}
\end{center}
\caption{\label{figure3} The real part of the correlator (\ref{avCC}) multiplied by $(-1)$
as a function of input signal power $|C_m|^2$ for $f(t)$ from Eq.~(\ref{eqInitCond}). The noise power parameter is $Q=10^{-21}$~W/(km$\times$Hz).
Dashed doted, dashed, and solid lines correspond to analytic representation (\ref{avCC}) with NLO-corrections (\ref{avCCNLO2}) for time grid spacings $\Delta_1$, $\Delta_2$, $\Delta_3$, respectively. Circles, squares, and diamonds correspond to numerical results for time grid spacings $\Delta_1$, $\Delta_2$, $\Delta_3$, respectively.}
\end{figure}
%========FIGURE-4
\begin{figure}[h]
\begin{center}
\includegraphics[width=6.9cm]{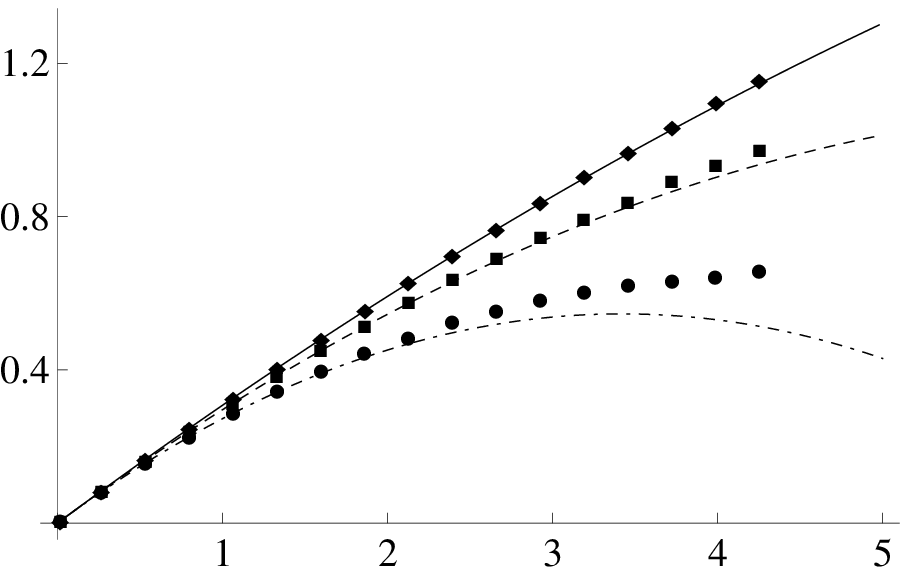}
\begin{picture}(0,0)
\put(-96,-10){\text{$|C_m|^2$, \,[mW]}}
\put(-220, 0){\rotatebox{90}{$-10 \mathrm{Im}\langle \left(\widetilde{C}_m - \langle \widetilde{C}_m \rangle \right)^2 \rangle   , \,[\mu W]$}}
\end{picture}
\end{center}
\caption{\label{figure4} The imaginary part of the correlator (\ref{avCC}) multiplied by $(-10)$
as a function of input signal power $|C_m|^2$ for $f(t)$ from Eq.~(\ref{eqInitCond}). The noise power parameter is $Q=10^{-21}$~W/(km$\times$Hz).
Dashed doted, dashed, and solid lines correspond to analytic representation (\ref{avCC}) with NLO-corrections (\ref{avCCNLO2}) for time grid spacings $\Delta_1$, $\Delta_2$, $\Delta_3$, respectively. Circles, squares, and diamonds correspond to numerical results for time grid spacings $\Delta_1$, $\Delta_2$, $\Delta_3$, respectively.}
\end{figure}
%========FIGURE-5
\begin{figure}[h]
\begin{center}
\includegraphics[width=6.9cm]{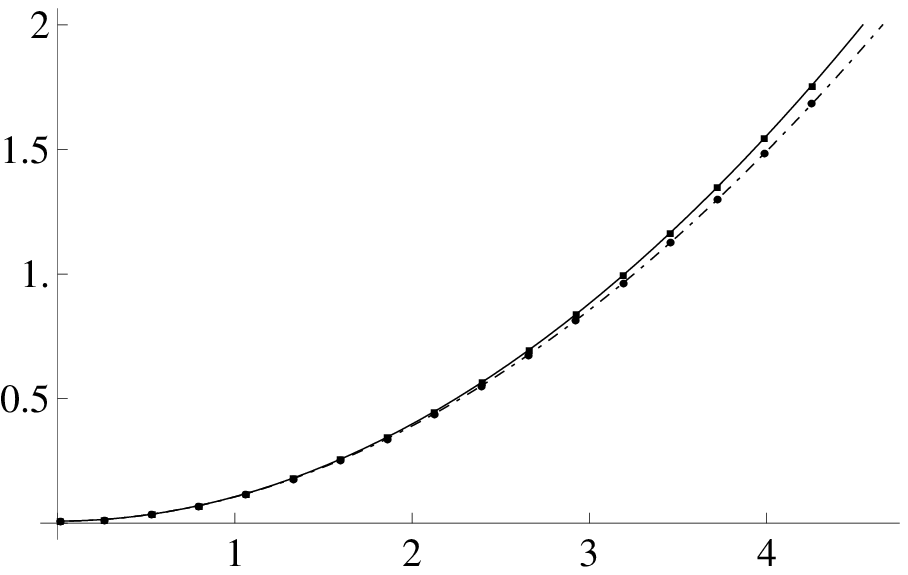}
\begin{picture}(0,0)
\put(-96,-10){\text{$|C_m|^2$, \,[mW]}}
\put(-220,19){\rotatebox{90}{$\langle \left|\widetilde{C}_m - \langle \widetilde{C}_m \rangle \right|^2 \rangle   , \,[\mu W]$}}
\end{picture}
\end{center}
\caption{\label{figure5} The  correlator (\ref{avCbarC}) as a
function of input signal power $|C_m|^2$ for $f(t)$ from Eq.~(\ref{eqInitCond}). The noise power parameter is $Q=10^{-21}$~W/(km$\times$Hz).
Dashed doted, dashed, and solid lines correspond to analytic representation (\ref{avCbarC}) with NLO-corrections (\ref{avCbarCNLO1}) for time grid spacings $\Delta_1$,  $\Delta_3$, respectively. Circles and squares  correspond to numerical results for time grid spacings $\Delta_1$, $\Delta_3$, respectively.}
\end{figure}
In Figs.~\ref{figure1}--\ref{figure5}  the numerical and analytical
results for correlators (\ref{avC})--(\ref{avCbarC}) are presented
for different time grid spacing $\Delta$
 {as a function of input signal
power}. In Fig.~\ref{figure5} the results are presented for the grid
spacings $\Delta_1$ and $\Delta_3$ because the results for
$\Delta_1$ and $\Delta_2$ almost coincide. One can see that
numerical and analytical results are in a good agreement
 {up to 3~mW at least}. However the
difference between numerical and analytical results for the smallest
time grid spacing $\Delta_1$ is maximal. Decreasing of the parameter
$\Delta$ means the increasing of the spectral bandwidth of the
noise. This increasing results in the growth of the total noise
power received by detector. Note that the analytical expressions for
correlators were obtained using the conditional PDF $P[Y(t)|X(t)]$
in the form (\ref{PYXapprox}). This form was derived in the
approximation of large signal-to-noise ratio
$\mathrm{SNR}=P\Delta/(Q L)$. Decreasing parameter $\Delta$ we
diminish  the parameter $\mathrm{SNR}$ and, as a consequence, the
accuracy of our approximation. The difference between numerical and
analytical results can be explained by taking into account the
next-to-leading order (NLO) corrections in noise power parameter $Q
L/\Delta$.  {Analytical results in
Figs.~~\ref{figure3}--\ref{figure5} are shown  with taking into
account both leading order  results (\ref{avCC})--(\ref{avCbarC}) and
NLO corrections presented in Appendix
\ref{AppendixA}, see Eqs. (\ref{avCCNLO2}) and (\ref{avCbarCNLO1}).}

%=======================================================
\subsection{Runge-Kutta method}

For the equation (\ref{startingCannelEqt}) the time $t$ is the
incoming parameter. Thus the simulation consists in the solution of
ordinary differential  equation with various initial conditions
determined by the real pulse shape $f(t)$, the amplitude   $C_m$,
and independent random noise functions $\eta(z,t)$. In the second
method we used pulse envelopes of the form
\begin{eqnarray}\label{fn}
f_n(t)=A_n\cos^n(\pi t/T_0)
\end{eqnarray}
for $n=2,4$, $t\in [-T_0/2,\, T_0/2]$: $A_2=\sqrt{\frac{8}{3}}$ and
$A_4=\sqrt{\frac{128}{35}}$. We choose the time discretization parameter $\Delta={T_0}/{64}$. The
random noise was realized as the telegraph  process with the step of
the length $\Delta_z=10^{-4} L$ and of the random height with zero
average and with the dispersion $\sigma^2=2.38
\times10^{-8}~W/(\mathrm{km}^2)$  both for real and imaginary parts.
The noise power parameter reads as $Q=2 \sigma^2 \Delta \Delta_z
\approx 5.94\times10^{-21}$~W/(km$\times$Hz) and it is almost six
times greater than that in the previous method. We independently
control this parameter $Q$ by using the leading order contribution
to the correlator (\ref{avCbarC}) numerically simulated for
$\gamma=0$. The noise $\eta$ is constant within the step. Within the
step the equation (\ref{startingCannelEqt}) was solved by the
Runge-Kutta method of the fourth order with the step
$h=\Delta_z/50$. The recovered input signal $\widetilde{ X} (t_j)$
was calculated using Eq.~(\ref{tildeXt}) at the equidistant points
$t_j$. The coefficients $\widetilde{C_k}$ were calculated using
Eq.~(\ref{tildeCk}). The average (\ref{avC}) and correlators
(\ref{avCC}),(\ref{avCbarC}) were calculated over 16384 values of
$\widetilde{C_k}$ which were found for various noise realizations.

To control the accuracy  of the method we solved the equation
(\ref{startingCannelEqt}) with zero noise from $z=0$ to $z=L$ with
the step $h$ and then we performed the backward propagation from
$z=L$ to $z=0$ with the found solution as the initial condition. In
the procedure the input signal was recovered with the relative
precision equal to $10^{-6}$.

\begin{figure}[h]
\begin{center}
\includegraphics[width=6.9cm]{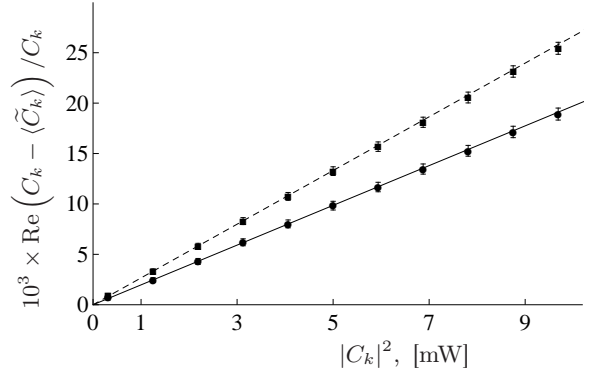}
\begin{picture}(0,0)
\put(-96,-10){\text{$|C_k|^2$, \,[mW]}}
\put(-220,10){\rotatebox{90}{$10^3\times \mathrm{Re} \left({C}_k-\langle \widetilde{C}_k \rangle  \right)/{C}_k $}}
\end{picture}
\end{center}
\caption{\label{figure6} The real part of the relative difference of  the coefficient $C_k$ and the correlator (\ref{avC}) in units $10^{-3}$
as a function of input signal power $|C_k|^2$ for $f_2(t)$, see black solid line, and for $f_4(t)$, see black dashed line. The noise power parameter is $Q=5.94\times10^{-21}$~W/(km$\times$Hz).  Circles and rectangles correspond to numerical results with statistic error on the level of three standard deviations for the functions $f_2$ and $f_4$, respectively.}
\end{figure}

\begin{figure}[h]
\begin{center}
\includegraphics[width=6.9cm]{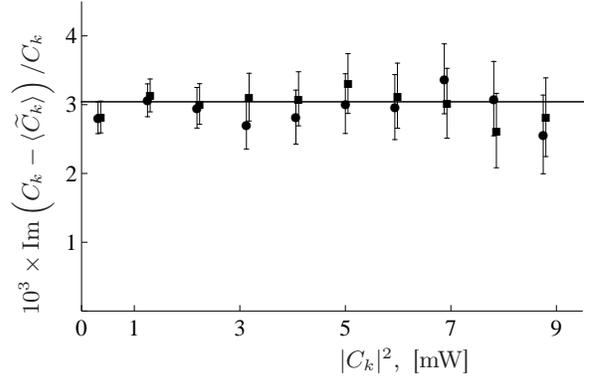}
\begin{picture}(0,0)
\put(-96,-10){\text{$|C_k|^2$, \,[mW]}}
\put(-220,10){\rotatebox{90}{$10^3\times \mathrm{Im} \left({C}_k-\langle \widetilde{C}_k \rangle  \right)/{C}_k $}}
\end{picture}
\end{center}
\caption{\label{figure7} The imaginary part of the relative difference of  the coefficient $C_k$ and the correlator (\ref{avC}) in units $10^{-3}$
as a function of input signal power $|C_k|^2$ for $f_2(t)$ and  $f_4(t)$, see black solid line. The noise power parameter is $Q=5.94\times10^{-21}$~W/(km$\times$Hz). Circles and rectangles correspond to numerical results with statistic error on the level of three standard deviations for the functions $f_2$ and $f_4$, respectively.}
\end{figure}

\begin{figure}[h]
\begin{center}
\includegraphics[width=6.9cm]{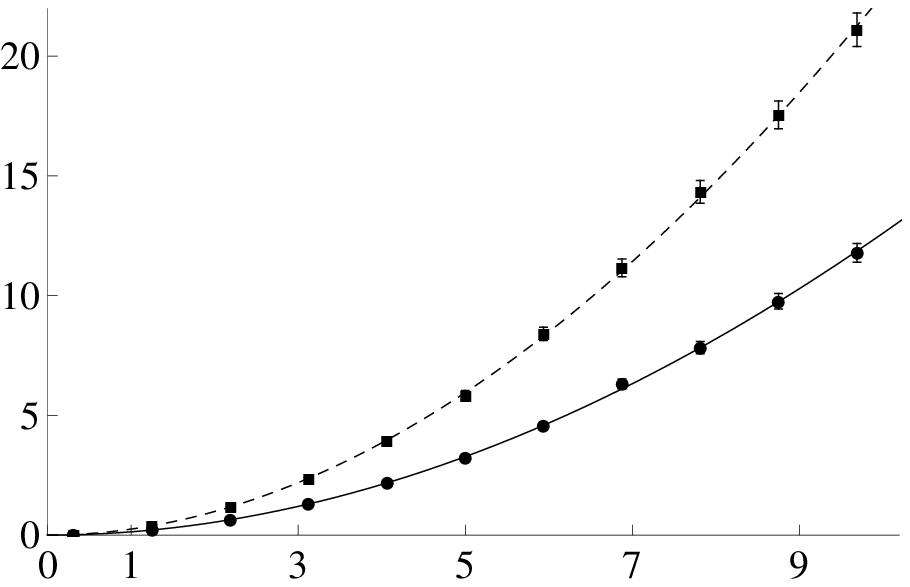}
\begin{picture}(0,0)
\put(-96,-10){\text{$|C_m|^2$, \,[mW]}}
\put(-220,10){\rotatebox{90}{$-\mathrm{Re}\langle \left(\widetilde{C}_m - \langle \widetilde{C}_m \rangle \right)^2 \rangle, \,[\mu W]$}}
\end{picture}
\end{center}
\caption{\label{figure8} The real part of the correlator (\ref{avCC}) multiplied by $(-1)$
as a function of input signal power $|C_m|^2$  for $f_2(t)$, see black solid line, and for $f_4(t)$, see black dashed line. Solid and dashed lines correspond to the real part of leading order contribution (\ref{avCC}) with the next-to-leading order corrections, see Eq.~(\ref{avCCNLO2}). The noise power parameter is $Q=5.94\times10^{-21}$~W/(km$\times$Hz).  Circles and rectangles correspond to numerical results with statistic error on the level of three standard deviations for the functions $f_2$ and $f_4$, respectively.}
\end{figure}

\begin{figure}[h]
\begin{center}
\includegraphics[width=6.9cm]{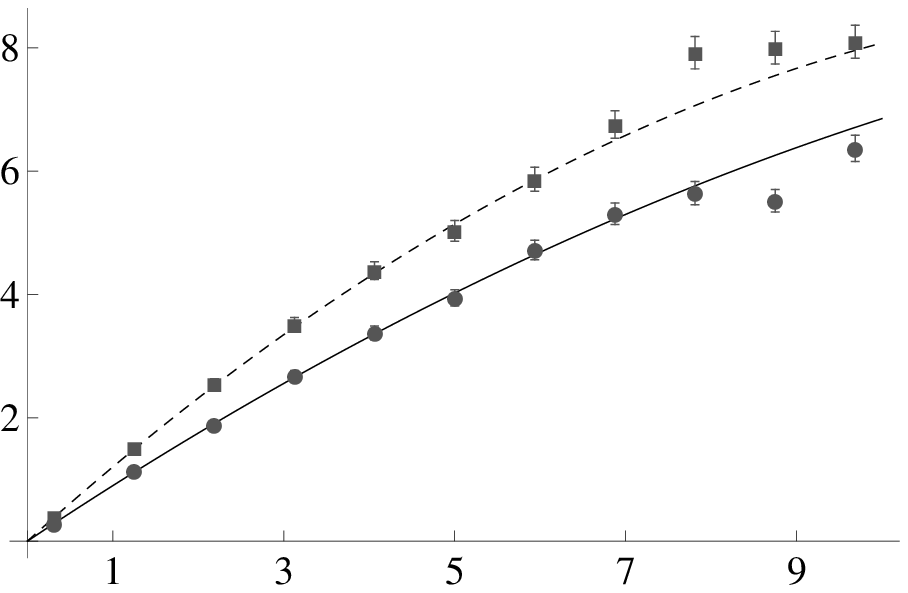}
\begin{picture}(0,0)
\put(-96,-10){\text{$|C_m|^2$, \,[mW]}}
\put(-220,10){\rotatebox{90}{$-10\mathrm{Im}\langle \left(\widetilde{C}_m - \langle \widetilde{C}_m \rangle \right)^2 \rangle   , \,[\mu W]$}}
\end{picture}
\end{center}
\caption{\label{figure9} The imaginary part of the correlator (\ref{avCC}) multiplied by $(-10)$
as a function of input signal power $|C_m|^2$ for $f_2(t)$, see black solid line, and for $f_4(t)$, see black dashed line. Solid and dashed lines correspond to the imaginary part of leading order contribution (\ref{avCC}) with the next-to-leading order corrections, see Eq.~(\ref{avCCNLO2}). The noise power parameter is $Q=5.94\times10^{-21}$~W/(km$\times$Hz).  Circles and rectangles correspond to numerical results with statistic error on the level of three standard deviations for the functions $f_2$ and $f_4$, respectively.}
\end{figure}

\begin{figure}[h]
\begin{center}
\includegraphics[width=6.9cm]{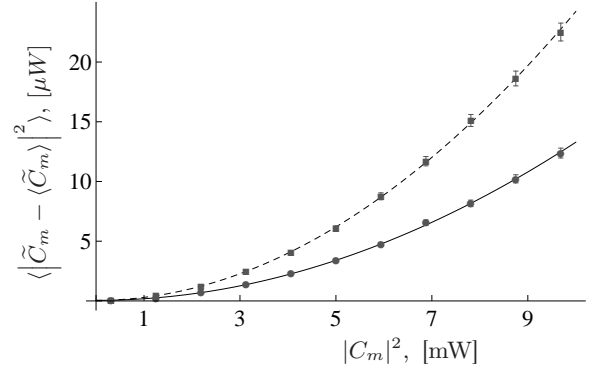}
\begin{picture}(0,0)
\put(-96,-10){\text{$|C_m|^2$, \,[mW]}}
\put(-220,19){\rotatebox{90}{$\langle \left|\widetilde{C}_m - \langle \widetilde{C}_m \rangle \right|^2 \rangle   , \,[\mu W]$}}
\end{picture}
\end{center}
\caption{\label{figure10} Correlator (\ref{avCbarC}) as a
function of input signal power $|C_m|^2$ for $f_2(t)$, see black solid line, and for $f_4(t)$, see black dashed line. Solid and dashed lines correspond to leading order contribution (\ref{avCbarC}) with the next-to-leading order corrections, see Eq.~(\ref{avCbarCNLO1}). The noise power parameter is $Q=5.94\times10^{-21}$~W/(km$\times$Hz).  Circles and rectangles correspond to numerical results with statistic error on the level of three standard deviations for the functions $f_2$ and $f_4$, respectively.}
\end{figure}

The  analytical  results in comparison with the numerical results  are presented in Figs.~\ref{figure6}--\ref{figure10} for different pulse shapes and average power. The numerical results are presented with  statistic errors on the level of three standard deviations.  One can see that numerical and analytical results are in a good agreement as well.
%========FIGURE-6
%========FIGURE-7
%========FIGURE-8
%========FIGURE-9
%========FIGURE-10

%==========================================Mutual information
\section{Entropies and mutual information}
\label{Section5}
\begin{figure}[t]
\begin{center}
\includegraphics[width=7cm]{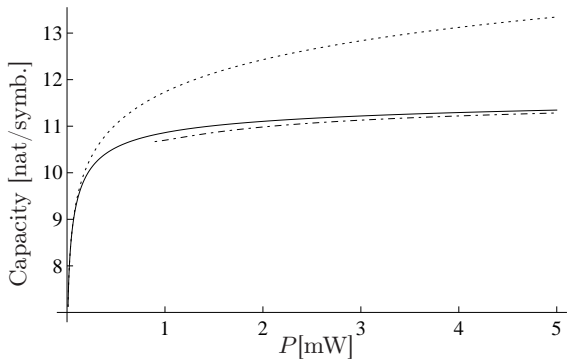}
\begin{picture}(0,0)
\put(-113,-5){\text{$P$[mW]}} \put(-215,20){\rotatebox{90}{\text{Capacity
[nat/symb.]}}}
\end{picture}
\end{center}
\caption{\label{figure11} Shannon capacity and the mutual information
$I_{P^{(m)}_{opt}}$ for the parameters $Q=10^{-21}$~W/(km$\times$Hz);  $L=800$~km;  $\gamma=1.25$~(km$\times$W)$^{-1}$; $T_0= 10^{-10}$~sec, and for the Gaussian shape (\ref{eqInitCond}) of $f(t)$. The black dotted line corresponds to the Shannon limit
$\log\left(\frac{P T_0}{Q L}\right)$, the black solid line corresponds to $I_{P^{(m)}_{opt}}$, see
Eq.~(\ref{Ipopt}), the black dashed dotted line corresponds to the asymptotics (\ref{optimalMIlarge}) for large $\gamma L P$. }
\end{figure}

 Now we proceed to the calculation of the output signal entropy
\begin{eqnarray}\label{HtildeC}
H[\widetilde{C}_m]= -\int d^2 \widetilde{C}_m P^{(m)}_{out}[\widetilde{C}_m] \log P^{(m)}_{out}[\widetilde{C}_m],
\end{eqnarray}
conditional entropy
\begin{eqnarray}\label{HCtildeC}
 H[\widetilde{C}_m|{C}_m] &=& -\int d^2 \widetilde{C}_m d^2 {C}_m P_{m}[\widetilde{C}_m|C_m] \times \nonumber \\&& P^{(m)}_{X}[C_m]  \log P_{m}[\widetilde{C}_m|C_m],
\end{eqnarray}
and the mutual information
\begin{eqnarray}\label{mutinf}
I_{P^{(m)}_X}=H[\widetilde{C}_m]- H[\widetilde{C}_m|{C}_m].
\end{eqnarray}
Our calculations of the entropies (\ref{HtildeC}), (\ref{HCtildeC}), and the mutual information (\ref{mutinf})  are similar to calculations of the entropies and the mutual information for per-sample channel, see Sec.III and Sec.IV of Ref.~\cite{Terekhov:2016a}.
Therefore we will not repeat the similar calculations here and present only the final results:
\begin{eqnarray}\label{HtildeCresult}
H[\widetilde{C}_m]&=& H[{C}_m]=\nonumber \\&&-\int d^2 {C}_m  P^{(m)}_{X}[C_m] \log  P^{(m)}_{X}[C_m],
\end{eqnarray}
\begin{eqnarray}\label{HCtildeCresult}
&&H[\widetilde{C}_m|{C}_m]=1+\log\left[\pi\frac{{ Q L}}{T_0}\right]+ \nonumber \\&& \frac{1}{2} \int d^2 C_m P^{(m)}_{X}[C_m] \log\left[1+\xi^2\frac{\gamma^2
L^2 |C_m|^4}{3} \right].
\end{eqnarray}
To calculate the optimal input signal distribution $P^{(m)}_{opt}[C_m]$ we calculate the mutual information substituting Eqs.~(\ref{HtildeCresult}) and (\ref{HCtildeCresult}) to Eq.~ (\ref{mutinf}) then we variate the mutual information over $P^{(m)}_{X}[C_m]$ with taking into account the normalization condition (\ref{normalization0}) and the fixed average power (\ref{avpower-1}). 
Assuming the variation of the mutual information to be zero, we obtain the equation for the optimal input signal distribution $P^{(m)}_{opt}[C_m]$. We solve the equation and obtain (for details of the similar calculations for per-sample channel see the Sec.III of Ref.~\cite{Terekhov:2016a}):
\begin{eqnarray}\label{Popt}
P^{(m)}_{opt}[C_m]=N_0 \frac{e^{-\lambda_0 |C_m|^2}}{\sqrt{1+\xi^2{\gamma^2
L^2 |C_m|^4}/{3}}},
\end{eqnarray}
where parameters $N_0=N_0 (P,\xi \gamma)$ and $\lambda_0=\lambda_0(P,\xi \gamma)$ are functions of the power $P$ and modified nonlinearity parameter $\xi \gamma $ by virtue of the relations (compare with Eqs. (46) and (47) of Ref.~\cite{Terekhov:2016a}):
\begin{eqnarray}\label{normalization1}
\int d^2 C_m P^{(m)}_{opt}[C_m]= \int^{\infty}_{0}d\rho \frac{ 2\pi N_{0} \,  \rho \,
e^{- \lambda_{0} \rho^2}}{\sqrt{1+ \xi^2 \gamma^2 L^2 \rho^4/3}} =1,
\end{eqnarray}
\begin{eqnarray}\label{normalization2}
&&P=\int \!\!  d^2 C_m P^{(m)}_{opt}[C_m] |C_m|^2  = \nonumber \\&&
\int^{\infty}_{0} \!\! d\rho\frac{ 2\pi N_{0}  \, \rho^3 e^{- \lambda_{0}
\rho^2}}{\sqrt{1+\xi^2 \gamma^2 L^2 \rho^4/3}} .
\end{eqnarray}
The capacity of one channel $m$, i.e., the mutual information calculated using the optimal input signal distribution (\ref{Popt}) reads
\begin{eqnarray}\label{Ipopt}
C=I_{P^{(m)}_{opt}}&=&\log\left( \frac{P T_0}{\pi e Q L}\right)+  P \lambda_0 - \log\left[P N_0\right].
\end{eqnarray}
One can see that the first term in the right-hand side of Eq.~(\ref{Ipopt}) corresponds to the Shannon's result \cite{Shannon:1948} for the linear channel at large signal-to-noise ratio, the second and third terms are related with the nonlinearity impact.  The result (\ref{Ipopt}) is similar to that obtained for the per-sample model in Ref.~\cite{Terekhov:2016a} but with modification of the Kerr nonlinearity parameter $\gamma$ for  per-sample model to  parameter $\xi \gamma$ for the present model, where $\xi=\sqrt{4n_6 - 3 n^2_4}$. There is no simple analytical form for $N_{0}$ and $\lambda_{0}$, see the Secs. III and IV of Ref.~\cite{Terekhov:2016a}, therefore we present below the analytical results for the asymptotics of the mutual information for small and large dimensionless nonlinearity parameter $\xi \gamma L P$ and the numerical calculations in Fig.~\ref{figure11}.
%=================FIGURE 11

Performing the substitution  $\gamma \rightarrow \xi \gamma$ in the
results of the Sec.III and Sec IV of Ref.~\cite{Terekhov:2016a} we
arrive at following asymptotics of the mutual information for small
and large dimensionless nonlinearity parameter $\gamma L P$:
\begin{eqnarray}\label{optimalMIsmall}
I_{P_X^{opt}[X]} \approx \log\left(\frac{P T_0}{Q L}\right)-\frac{\xi^2 \gamma^2 L^2 P^2}{3},
\end{eqnarray}
for $\xi \gamma L P \ll 1$, and
\begin{eqnarray}\label{optimalMIlarge}
&&\!\!\!I_{P_X^{opt}[X]}= \log\log\left({B \xi \gamma L
P}/{\sqrt{3}}\right)-\log\left({{QL^2\xi \gamma e}}/{\sqrt{3}}\right)+\nonumber \\&&
\frac{1}{\log\left({B \xi \gamma L P}/{\sqrt{3}}\right)}\Bigg[ \log\log\left({B \xi \gamma L
P}/{\sqrt{3}}\right)+1 - \nonumber \\&& \frac{\log\log\left({B \xi \gamma L
P}/{\sqrt{3}}\right)}{\log\left({B \xi \gamma L P}/{\sqrt{3}}\right)} \Bigg],
\end{eqnarray}
for $  \log \xi \gamma L P \gg 1$ and $P \ll \Delta/(Q L^3 \xi^2 \gamma^2)$. Here $B= 2  e^{-\gamma_{E}}$, $\gamma_{E} \approx 0.5772$ is the Euler constant. Note that the asymptotics (\ref{optimalMIlarge}) is obtained with accuracy $1/\log^2(\xi \gamma L P)$, see the Sec. IV of Ref.~\cite{Terekhov:2016a}.

%==========================================Conclusion
\section{Conclusion}

In the present paper we use  results obtained in
Ref.~\cite{Terekhov:2016a} for per-sample model to calculate the
informational characteristics of the channel where the input signal
$X(t)$ depends on time, see Eq.~(\ref{Xtmodelg}). For this channel
the information is carried by coefficients $C_k$. In the process of
the signal propagation the input signal is transformed by the Kerr
nonlinearity and the noise in the channel. To recover the transmitted
information we introduce the detection procedure which removes the
nonlinearity effects, see Eq.~(\ref{tildeXt}), and then projects
$\widetilde{X}(t)$ on the basis functions, see Eq.~(\ref{tildeCk}),
to obtain the coefficients $\widetilde{C}_k$. Using the conditional
probability density function for per-sample model obtained in
Ref.~\cite{Terekhov:2016a} we calculate the correlators of the
coefficients $\widetilde{C}_k$, see
Eqs.~(\ref{avC})--(\ref{avCbarC}). We demonstrate that these
correlators depend on the noise bandwidth parameter $\Delta$.
We also perform the numerical calculations of these correlators
using two different methods and show that the numerical and
analytical results are in agreement. Using obtained results for
correlators we find the conditional probability density function
$P[\{\widetilde{C}_k\}|\{{C}_k\}]$ in the leading and
next-to-leading orders in parameter $Q L/(\Delta P)$. Then we
calculate the informational entropies and the mutual information for
the channel in leading order in the parameter $Q L/(T_0 P)$. We
perform variation of the mutual information over the input signal distribution
function and obtain the optimal input signal distribution function
which maximizes the mutual information. We calculate the channel
capacity in the leading order in parameter $Q L/(T_0 P)$ and
demonstrate that the capacity depends on the pulse envelope  through
one parameter $\xi$, see Eq.~(\ref{xi}). The capacity grows as $\log \log P$ for
sufficiently large average power $P $: $ (\xi \gamma Q L )^{-1} \ll
P \ll \Delta/(Q L^3 \xi^2 \gamma^2)$. Note that the same asymptotics
was obtained for per-sample model, therefore taking into account the
time dependance of the pulse envelope does not change the
asymptotics behavior and modifies only the nonlinearity parameter
$\gamma$ to $\xi \gamma$.

%==========================================Acknowledgments (правильно напиши!)
\begin{acknowledgments}
\emph{Acknowledgment}\\
All authors would like to thank the Russian Science Foundation (RSF), grant No.
16-11-10133. Also A.V. would like to thank the Russian Foundation for Basic Research (RFBR), grant No. 16-31-60031. Also I.S. would like to thank the Russian Science Foundation (RSF), grant No. 17-72-30006, and Ministry of Education and Science of the Russian Federation (14.Y26.31.0017).
\end{acknowledgments}

%==========================================Appendix
\appendix

\section{Correlators (\ref{avCC}) and (\ref{avCbarC}) with NLO corrections \label{AppendixA}}

Let us present the correlator (\ref{avCC}) with next-to-leading (NLO) corrections in the noise power.
\begin{eqnarray}\label{avCCNLO2}
&&\!\!\!\!\!\left\langle \left(\widetilde{C}_m - \langle \widetilde{C}_m \rangle \right)\left(\widetilde{C}_n - \langle \widetilde{C}_n \rangle \right)  \right\rangle= \delta_{m,n} \Bigg( \nonumber \\&&  \left\langle \left(\widetilde{C}_m - {C}_m \right)\left(\widetilde{C}_m - {C}_m \right)  \right\rangle - \left(\frac{ Q L^2 \gamma}{\Delta}\right)^2 {C}_m^2 \Big\{ \nonumber \\&& -1 + \frac{n^2_4}{9}\gamma^2 L^2 |{C}_m|^4 + i \frac{2 n_4}{3} \gamma L  |{C}_m|^2\Big\}\Bigg)= \nonumber \\&&
\delta_{m,n}\Bigg(  \frac{   {Q} L^2 \gamma}{T_0}{C}^2_m \left[-\frac{2 n_6}{3}\gamma L |{C}_m|^2 -i n_4 \right]+\nonumber \\&& \left(\frac{   {Q} L^2 \gamma}{T_0}\right)^2 \frac{T_0}{\Delta}{C}^2_m \Bigg[-\frac{9 n_4}{2} + \frac{2 n_8}{3} \gamma^2 L^2 |{C}_m|^4 + \nonumber \\&& i\frac{58 n_6 }{15}   \gamma L |{C}_m|^2 \Bigg]\Bigg).
\end{eqnarray}
Here we have used the relation
\begin{eqnarray}
&&\!\!\!\!\!\left\langle \left(\widetilde{C}_m - \langle \widetilde{C}_m \rangle \right)\left(\widetilde{C}_m - \langle \widetilde{C}_m \rangle \right)  \right\rangle= \nonumber \\&& \!\!\!\!\!\left\langle \left(\widetilde{C}_m - {C}_m \right)\left(\widetilde{C}_m - {C}_m \right)  \right\rangle - \left\langle \widetilde{C}_m - {C}_m \right \rangle^2,
\end{eqnarray}
the result (\ref{avC}) for $ \left\langle \widetilde{C}_m - {C}_m
\right \rangle$ and the calculation of $\left\langle
\left(\widetilde{C}_m - {C}_m \right)\left(\widetilde{C}_m - {C}_m
\right)  \right\rangle$ on the base of next-to-leading order result
for $P[Y|X]$ in Ref.~\cite{Terekhov:2017}.

\begin{figure}[t]
\begin{center}
\includegraphics[width=6.9cm]{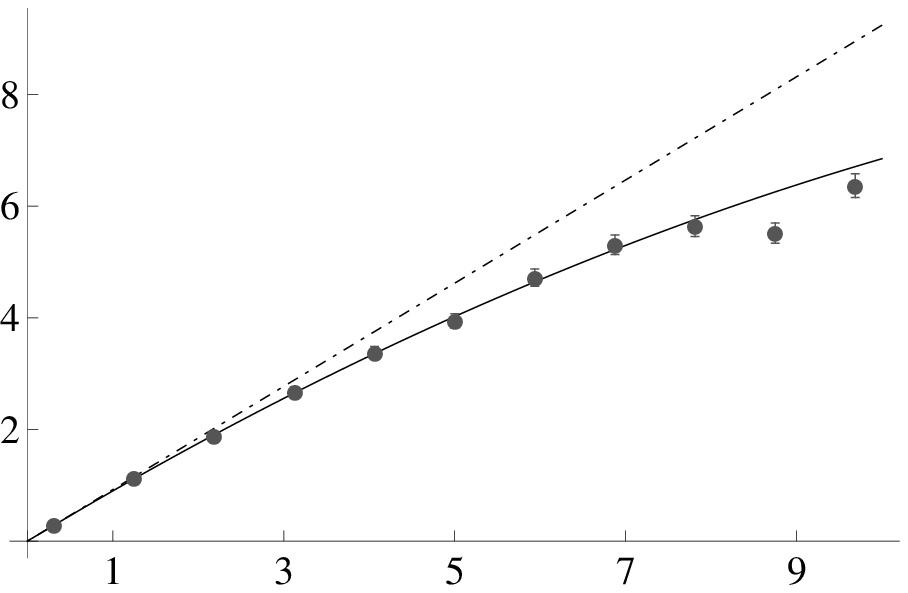}
\begin{picture}(0,0)
\put(-96,-10){\text{$|C_m|^2$, \,[mW]}}
\put(-220,0){\rotatebox{90}{$-10\mathrm{Im}\langle \left(\widetilde{C}_m - \langle \widetilde{C}_m \rangle \right)^2 \rangle   , \,[\mu W]$}}
\end{picture}
\end{center}
\caption{\label{figure12} The imaginary part of the correlator (\ref{avCC}) multiplied by $(-10)$
as a function of input signal power $|C_m|^2$ for $f_2(t)=\sqrt{\frac{8}{3}}\cos^2(\pi t/T_0)$ in the leading order (\ref{avCC}), see black dashed dotted line, and with the next-to-leading order corrections (\ref{avCCNLO2}), see the solid line.  The noise power parameter $Q=5.94\times10^{-21}$~W/(km$\times$Hz). Circles represent the numerical results for Runge-Kutta method.}
\end{figure}

In a similar manner it is easy to  calculate  the following corrections to
correlator (\ref{avCbarC}) from the results
obtained in Ref.~\cite{Terekhov:2017}:
\begin{eqnarray}\label{avCbarCNLO1}
&&\left\langle \left(\widetilde{C}_m - \langle \widetilde{C}_m \rangle \right)\overline{\left(\widetilde{C}_n - \langle \widetilde{C}_n \rangle \right)}  \right\rangle= \delta_{m,n}\Bigg( \nonumber \\&& \left\langle \left(\widetilde{C}_m - {C}_m \right)\overline{\left(\widetilde{C}_m - {C}_m \right)}  \right\rangle- \nonumber \\&& \left(\frac{ Q L^2 \gamma}{\Delta}\right)^2 |{C}_m|^2 \Bigg\{1+\frac{n^2_4}{9}\gamma^2 L^2 |{C}_m|^4\Bigg\}\Bigg)=\nonumber \\&& \delta_{m,n}\Bigg(
 \frac{{Q} L}{T_0} \left[1+\frac{2 n_6}{3}\gamma^2 L^2 |{C}_m|^4\right]+ \nonumber \\&&  \left(\frac{   {Q} L^2 \gamma}{T_0}\right)^2 \frac{T_0}{\Delta}|{C}_m|^2\Bigg[ n_4 -\frac{2 n_8}{9}\gamma^2 L^2 |{C}_m|^4 \Bigg] \Bigg).
\end{eqnarray}
Note that these NLO results  (\ref{avCCNLO2}) and (\ref{avCbarCNLO1})
contain  the time discretization parameter $\Delta$ related with the noise bandwidth $W'=2\pi/\Delta$. The relative
importance of the NLO corrections in correlators (\ref{avCCNLO2})
and (\ref{avCbarCNLO1}) is governed by the dimensionless parameter
$\left(\frac{Q L}{\Delta}\gamma L\right) \gamma L P $, i.e., it
increases linearly for large and increasing $P $. To demonstrate the
importance of these corrections for our numerical results we present
the Fig.~(\ref{figure12}) where for the noise power parameter
$Q=5.94\times10^{-21}$~W/(km$\times$Hz) the imaginary part of the
leading order contribution (\ref{avCC}) and the next-to-leading
order corrections (\ref{avCCNLO2}) are presented together with the
numerical results (Runge-Kutta method) for the envelope form
$f_2(t)=\sqrt{\frac{8}{3}}\cos^2(\pi t/T_0)$.
{One can see that our calculations,
i.e., Eq.~(\ref{PmCC})  and formulae of the Sec. \ref{Section5}
based on the leading order results (\ref{avC})--(\ref{avCbarC}) are
in a good agreement  with the numerical calculations up to the
average power of order of 4~mW for given noise and channel
parameters.}
%========FIGURE-12

%===========================================Bibliography in PRE style

\end{document}